\documentstyle[prl,aps]{revtex}
\input{psfig}
\draft
\begin{document}

\twocolumn[\hsize\textwidth\columnwidth\hsize\csname@twocolumnfalse\endcsname
\title{Hexagons  and Interfaces in a Vibrated Granular Layer}

\author{I.S.  Aranson $^{1}$, L.S.  Tsimring$^2$,  and V.M.  Vinokur $^1$}
\address{
$^1$ Argonne National Laboratory,
9700 South  Cass Avenue, Argonne, IL 60439 \\
$^2$Institute for Nonlinear Science, University of California, San Diego, La Jolla,
CA 92093-0402
}
\date{\today}
\maketitle 

\begin{abstract} 
The order parameter model based on parametric Ginzburg-Landau equation
 is used to describe high acceleration 
patterns in vibrated layer of granular material.  At large amplitude of driving both
hexagons and interfaces emerge. Transverse  instability  
leading to formation of ``decorated''  interfaces and labyrinthine patterns, 
is found.
Additional sub-harmonic forcing leads to controlled interface motion.   
\end{abstract} 

\pacs{PACS: 47.54.+r, 47.35.+i,46.10.+z,83.70.Fn}

\narrowtext
\vskip1pc]
Driven granular materials often manifest collective 
fluid-like behavior:  
convection, surface waves, and pattern formation (see e.g.
\cite{jnb}).  
One of the most fascinating  examples of  this collective dynamics is
the appearance of long-range coherent patterns and localized excitations
in vertically-vibrated thin granular layers
\cite{swin1,swin2,jag3}.
The particular pattern is determined by the interplay between driving frequency 
 $f$ and acceleration of the container $\Gamma= 4 \pi^2 {\cal A }  f^2/g$  
(${\cal A} $ is the amplitude of oscillations, $g$ is the gravity acceleration)  \cite{swin1,swin2}. 
Patterns appear  at   $\Gamma\approx 2.4$  almost
independently of the driving frequency $f$. 
At small frequencies $f< f^* $ \cite{herz}  the
transition is subcritical (hysteretic), leading to formation of  either squares or localized
excitations  such as individual {\it oscillons} and various bound states of
oscillons.   For  higher frequencies $f >f^* $ the transition is
supercritical, and stripes appear first. Both squares and stripes
oscillate at half of the driving frequency $f/2$. At higher acceleration
($\Gamma>4 $), stripes and squares become unstable, and hexagons appear
instead. Further increase of acceleration at $\Gamma \approx 4.5$ converts 
hexagons into a domain-like structure of flat layers oscillating with 
frequency $f/2$ with opposite phases. Depending on parameters, interfaces 
separating flat domains, are either smooth or ''decorated'' by
periodic undulations. Undulations slowly grow with time and eventually become
quasi-periodic labyrinthine patterns.  For $\Gamma >5.7$  various 
quarter-harmonic patters emerge. 

The problem of pattern formation in thin layers of granular material was 
studied by several groups. While molecular dynamics simulations 
\cite{luding} and hydrodynamic and phenomenological models \cite{rothman} 
reproduced  certain   experimental features, neither of them offered 
a systematic description of the whole rich variety of the observed phenomena.  
In Ref. \cite{ta97} we introduced the order parameter characterizing   
the complex amplitude of sub-harmonic oscillations. 
The equations of motion following from the symmetry arguments 
and mass conservation reproduced essential  phenomenology of
patterns near the threshold of primary bifurcation:  stripes, squares and 
oscillons. 

In this Letter  we describe high acceleration patterns on the basis the order 
parameter model. We  show  that at large amplitude of driving both
hexagons and interfaces emerge. We find morphological instability 
leading to formation of ``decorated''  interfaces and labyrinthine patterns.
We develop the description of the labyrinths 
in terms of nonlocal contour dynamics.
Additional sub-harmonic forcing leads to the motion of the interface with the 
direction controlled  
by  relative phases of harmonic and sub-harmonic components of forcing.  


The essential of the  model \cite{ta97,at98} is  the order parameter equation 
for the complex amplitude $\psi$ of parametric layer oscillations 
$h=\psi\exp(i\pi f t)+c.c.$ at the frequency
$f/2$. In general, $\psi$ is coupled with the the average thickness of the
layer $\rho$. However, at high frequencies ($f > f^* $) the coupling 
between  $\rho$ and  $\psi$ becomes unimportant, 
and the model can be reduced to a single order-parameter equation
\begin{eqnarray}
\partial_t\psi&=&\gamma\psi^*-(1-i\omega)\psi+(1+ib)\nabla^2\psi
-|\psi|^2\psi
\label{eq1}
\end{eqnarray}
Eq.(\ref{eq1}) describes the evolution of the
order parameter for the parametric instability in spatially-extended 
systems (see
\cite{vinals,kiyashko}). 
Linear terms in this equation  are obtained  from
the complex growth rate for infinitesimal periodic
layer perturbations $h \sim  \exp[ \Lambda(k) t + i k x]$. Expanding 
$\Lambda(k)$ for small $k$, and keeping only 
two leading terms in the expansion $\Lambda (k) = -\Lambda_0   -\Lambda_1 k^2$
we reproduce the linear terms in Eq. (\ref{eq1}), 
where $b = Im \Lambda _1/ Re \Lambda_1$  and $\omega = (\Omega_0- \pi f )
/Re \Lambda _0 $, where $\Omega_0 = - Im \Lambda_0$.   
The term  $\gamma \psi^*$ characterizes the effect of driving 
at the resonance  frequency. 
The term $-|\psi|^2\psi$ accounts for
nonlinear saturation of waves at finite amplitude.

It is convenient to shift  the phase of the complex order
parameter via $\tilde{\psi}=\psi\exp(i\phi )$ with $\sin
2\phi=\omega/\gamma$. 
The equations  for real and imaginary part 
$\tilde \psi = A + i B$ are:
\begin{eqnarray}
\partial_t A &=&(s -1) A -  2  \omega B - ( A^2+ B^2) A +\nabla^2 ( A- b B) 
\label{eq2a} \\ 
\partial_t B  &=& -  (s + 1)  B - ( A^2+ B^2) B +\nabla^2 ( B + b  A ) 
\label{eq2b}
\end{eqnarray}
where $s^2 = \gamma^2 -\omega^2$.
At $s<1$, Eqs.  (\ref{eq2a}),(\ref{eq2b})
has only one trivial uniform state $A=0,\ B=0$, 
At $s>1$, two new uniform states  appear, $A=A_0 = \pm \sqrt{s-1},\ B=0$. 
The onset of these states corresponds to the period doubling
of the layer flights sequence,  observed in experiments
\cite{swin1} and predicted by the simple  inelastic ball model
\cite{swin1,metha}. Signs $\pm$ reflect two relative phases of 
layer flights with respect to container  vibrations.

First we analyze the stability of the state $A=A_0,
B=0$ 
with respect to perturbations with wavenumber $k$,
$(A, B) = (A_0, 0 ) + (U_k, V_k )
\exp [\lambda(k) t +  i k x ].$
The uniform state loses its stability
with respect to periodic modulations with the wavenumber $k_c$
at $s < s_c$ (correspondingly, $\gamma< \gamma_c$), where 
\begin{eqnarray} 
s_c &=&\frac{ \sqrt{(1+\omega^2)(1+b^2)}-\omega+b}{2b}  
\label{sc}  \\
k_c^2 &=&-  \frac{2s-1-\omega b} { 1+b^2} 
\label{kc} 
\end{eqnarray} 
Small perturbations with all 
directions of the wavevector grow with the same rate. The resultant 
selected pattern is determined by the nonlinear competition between
the modes. In the presence of the reflection symmetry $\psi\to
-\psi$, quadratic nonlinearity is absent, and cubic nonlinearity favors
stripes corresponding to a single mode.
Near the fixed points $A=A_0, B=0$ the reflection symmetry for
perturbations $U \to-U,\ V \to -V$ is broken, and 
hexagons   emerge  at the threshold of instability.  
To clarify  this point 
we perform  weakly-nonlinear analysis  of Eqs. (\ref{eq2a}),(\ref{eq2b}) 
for  $s = s_c - \epsilon$, and $\epsilon \ll 1 $. 
At $\epsilon \to 0$, the variables $U$ and $V$ are related as 
in linear system: 
\begin{eqnarray} 
(U, V) & =&  ( 1, \eta) \Psi \;,\; 
\eta  =  \left[ 2(s_c -1) + k^2_c\right]/( b k^2_c-2\omega) 
\nonumber   \\
\Psi & = & \sum A_j \exp [ i {\bf k r } ] + c.c., | k| = k_c \label{eq11} 
\end{eqnarray}  
Corresponding  adjoint eigenvector is  
\begin{eqnarray} 
(U^+ , V^+ )  =  ( 1, \eta ^+ )  \;,\;
\eta^+   =  - \left[ 2(s_c -1) + k^2_c\right]/ b k^2_c \label{eq13} 
\end{eqnarray} 
Substituting Eq. (\ref{eq11}) into Eqs. (\ref{eq2a}),(\ref{eq2b})  and performing the 
orthogonalization, we obtain  equations for the slowly-varying
complex amplitudes 
$A_j$, $j=1,2,3$ (we assume only three waves with triangular symmetry, favored by quadratic nonlinearity) 
\begin{eqnarray} 
\partial_t  A_j &=&  2 \epsilon A_j + a_2 A_{j+1}^* A_{j-1} ^*   \nonumber \\
&-&a_3 \left ( | A_j|^2+ 2( |A_{j-1}| ^2+ |A_{j+1}|^2 ) \right ) A_j 
\label{eq14} 
\end{eqnarray} 
where the coefficients $a_2,a_3$ are
\begin{eqnarray} 
a_2  =  2 A_0 \left(2 + \frac{1+\eta^2}{1+ \eta \eta^+ } \right) \;,\;
a_3  =  3 ( 1+ \eta \eta^+) 
\label{coef} 
\end{eqnarray} 
Eqs. (\ref{eq14}) are well-studied (see \cite{ch}). There are three critical values of 
$\epsilon$: $ \epsilon_A=-a_2^2/40a_3,\ \epsilon_R=a_2^2/2a_3,$ and $\epsilon_B=2a_2^2/a_3$. 
The hexagons are stable for $\epsilon_A < \epsilon < \epsilon_B$, and 
the stripes are 
stable for $\epsilon > \epsilon_R$. 
Thus, near $s=s_c$ the model exhibit stable  hexagons \cite{belg}. 
Since we have two symmetric fixed points, 
both up- and down-hexagons co-exist. For smaller $s$
 stripes are stable, and for larger $s$, flat 
layers are stable, in agreement with observations\cite{swin2,blair}.
The above  analysis requires the values of $\epsilon_{A,B,R}$ to  be
small. For parameters $\omega, b= O(1)$, this requirement  is satisfied  for  $\epsilon_A  $, but not 
for  $\epsilon_{B,R}$. The estimates 
can be improved by substituting  $A_0$ at $s=s_c+\epsilon$ instead of $A_0(s_c)$  in 
(\ref{coef}). The resulting range of stable hexagons is plotted in the phase diagram $(\omega,\gamma)$
(see Fig. \ref{phase}). 

At $s>1$,  Eqs.    (\ref{eq2a}),(\ref{eq2b}) have an  interface solution connecting 
two uniform states
$A_0   = \pm \sqrt {s-1}  ,\ B=0$.
For $b=0$ this solution is of the form $ A= A_0 \tanh (A_0 x /2 ),
B=0$. For $b\ne 0$ the solution is available only numerically.
In order to investigate the stability of the interface, 
we consider the perturbed solution 
$(A, B) =  (A_0, B_0) + ( \tilde A (x), \tilde B(x))  \exp [\lambda(k) t + i k y]$. 
For $\tilde A , \tilde B$ we obtain linear equations
\begin{equation}
\hat L {\tilde A \choose  \tilde B}  =
(\lambda(k) + k^2) {\tilde A \choose  \tilde B} + b  k^2 {\tilde B
\choose -\tilde A}
\label{eq16}
\end{equation}
\begin{eqnarray}
\hat L =   \left (
\begin{array} {rl}
s -1 -3 A_0^2- B_0^2 +  \partial_x^2,   & -2  \omega
-2 A_0  B_0-b  \partial_x^2 \\
-2 A_0  B_0+b  \partial_x^2  ,  & -  s - 1-  A_0 ^2
- 3 B_0 ^2+ \partial_x^2
\end{array}
\right)
\nonumber 
\end{eqnarray}

\begin{figure}
\centerline{\psfig{figure=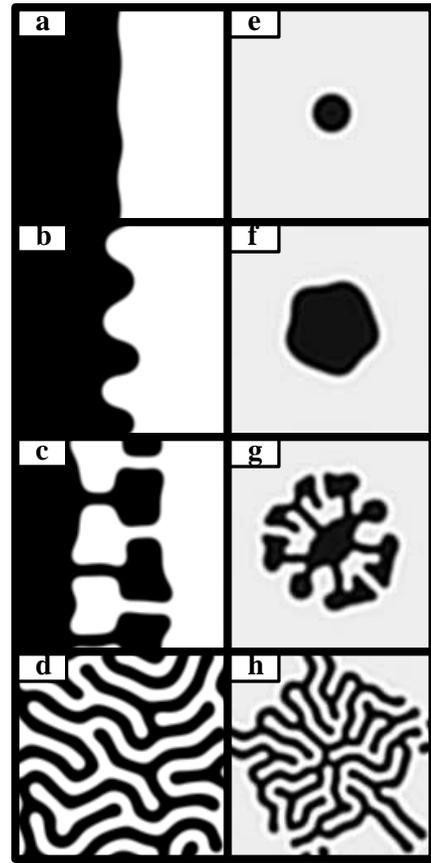,height=4.5in}}
\caption{ (a-d) Interface instability  and labyrinth formation,  
Eq. (\protect \ref{eq1}), $\omega=2, b=4, \gamma=2.9$, domain size 
$100\times 100$ units, the snapshot are taken at times $t=1000, 1600, 2300, 4640$; 
(e-h) labyrinth formation from a circular spot, SHE, $\delta=1.4$,  domain size 
$100 \times 100$ units, $t=200, 1300, 1700, 1900$.   
 }
\label{interf_fig}
\end{figure}

In order to determine the spectrum of eigenvalues $\lambda(k) $ one has
to solve Eq. (\ref{eq16}) along with stationary Eqs. (\ref{eq2a}),(\ref{eq2b}).  Using
numerical matching-shooting technique, we have obtained positive eigenvalues
corresponding to interface instability. This
instability is confirmed by direct numerical simulations of Eq.(\ref{eq1}). An
example of the evolution of slightly perturbed interface is shown in
Fig. \ref{interf_fig}. Small perturbations grow to form a ``decorated''
interface similar to \cite{umban}, with time these decorations evolve 
slowly and eventually form a labyrinthine pattern \cite{blair}.

The neutral curve for this instability can
be determined as follows. Numerical analysis shows that at the threshold
the most unstable wavenumber is $k=0$ and we can expect that for $k \to 0 $
$\lambda  \sim k^2$. Expanding Eqs.(\ref{eq16}) in power series
of $k^2$: $(\tilde A, \tilde B) = (A^{(0)}, B^{(0)})+
 k^2 (A^{(1)}, B^{(1)})+....$ in the zeroth order in $k$ we obtain
$\hat L (A^{(0)}, B^{(0})=0$.  The corresponding solution is the translation 
mode $A^{(0)}=\partial_x A_0(x), B^{(0)}= \partial_x B_0(x)$.
In the first order in $k^2$ we arrive at the linear inhomogeneous problem
\begin{equation}
\hat L { A^{(1)} \choose   B ^{(1)}}  =
(\lambda(k) + k^2) { A^{(0)} \choose   B^{(0)}} + b  k^2 { B^{(0)}
\choose - A^{(0)}}
\label{eq18}
\end{equation}
A bounded solution to Eq. (\ref{eq18}) exists if the r.h.s. is orthogonal to 
the localized mode of the adjoint operator $A^+, B^+$. The orthogonality 
condition fixes the relation between $\lambda $ and $k$: 
\begin{equation} 
\lambda = -(1+\beta ) k^2,
\beta = \frac{b\int_{-\infty }^\infty (A^{(0)} B^+ - B^{(0)} A^+) dx } 
{ \int_{-\infty }^\infty (A^{(0)} A^+ + B^{(0)} B^+) dx } 
\label{coef3}
\end{equation} 
The instability, corresponding to the negative surface 
tension of the interface,  onsets at $\beta = - 1$. 
The neutral curve is shown in Fig. \ref{phase}. This 
instability leads to  so called 
``decorated'' interfaces (see Fig. \ref{interf_fig},a). 
At the nonlinear stage 
the undulations grow and form labyrinthine patterns (Fig. \ref{interf_fig}, b-d). The  evolution of the
interface can be investigated near the line $s=1$  (see Fig.  \ref{phase}).
In the vicinity of this line  $ A \sim (s-1)^{1/2} $ and $ B \sim (s-1)^{3/2} \ll A$.
in the leading order we can obtain 
from Eq. (\ref{eq2b})  $B=b\nabla^2A/2$, and Eq. (\ref{eq2a}) 
yields
\begin{equation} 
\partial_t A =(s -1) A - A^3  + (1-\omega b) \nabla^2  A-\frac{ b^2}{2}  
 \ \nabla^4 A. 
\label{sh}
\end{equation} 
Rescaling the variables $t \to (s -1 ) t,$ 
$A \to (s-1)^{-1/2}A,$ $x \to (2(s-1)/b^2)^{1/4}$, we can reduce 
Eq. (\ref{eq1})  to the Swift-Hohenberg equation (SHE)  
\begin{equation}
\partial_t A = A - A^3  - \delta  \nabla^2  A-
 \ \nabla^4 A
\label{sh1}
\end{equation}
where 
$\delta = ( \omega b - 1 ) \sqrt {  2 /((s-1)b^2)}$. 
This equation is simpler than Eq. (\ref{eq1}), however
it captures many essential features of the original system dynamics,
including existence and stability of stripes and hexagons in
different parameter regions (see\cite{dewel}), 
existence of the interface solutions, interface instability and
emergence of labyrinthine patterns. Indeed, simple analysis shows that
the growth rate of the instability of the uniform state $A=1$ as a
function of the perturbation wavenumber is
determined by the formula $\lambda (k) = -2 + \delta k^2 -k^4$, so it
becomes unstable at  $\delta>\delta_c=2\sqrt{2}$ at critical wavenumber
$k_c=\sqrt{2}$. As in the original model, near the threshold of this
instability, subcritical hexagonal patterns are preferred. 
Interface stability can also be analyzed  more simply as the linearized 
operator corresponding to the model (\ref{sh1}) is self-adjoint. The
threshold value $\delta_{\mbox {th}}= 1.011$ is obtained from the following solvability
condition
\begin{equation} 
\int_{-\infty} ^ {\infty}  \left( \delta_{\mbox{th}} A_{0x}^2 - 2 A_{0xx}^2 \right)  d x
=0
\end{equation}
                            
We derived the equation for the position  of the curved interface  
${\bf R}$ parameterized by its arclength $\xi$.
The solution is sought for in the form 
$A = A_0({\bf n}\cdot ({\bf r} -{\bf R}(\xi)))+ W({\bf r})$,  
where ${\bf n}$ is the unit normal vector, $A_0$ is the 1D interface 
solution and $W$ is the correction 
due 
to interaction with the remote parts of the interface. This correction can 
be obtained from the linearized SHE in the far field.
Substituting this solution in the SHE (\ref{sh1}), we obtained 
the equation for the interface velocity $c_n={\bf n}\cdot \partial_t {\bf R }$,
\begin{eqnarray} 
 c_n=  
-\sigma \kappa + \frac{ 2 C \chi}{ \mu }  
 \oint d \xi^\prime  \zeta K_1( \chi |{\bf R}(\xi^\prime)-{\bf R}(\xi)|) 
   + c.c. 
\label{contour}  
\end{eqnarray} 
where $\kappa$ is the local interface curvature,  
$\mu = \int A_x^2 dx$ is an interface ``mass'' per unit length, and 
 $\sigma = - \delta + 2  \int A_{xx}^2 dx /\mu$ is the surface tension,  $ \zeta = 
({\bf R}(\xi) -{\bf R}(\xi^\prime))/| {\bf R}(\xi)-{\bf R}(\xi^\prime)| 
\times {\bf R}_\xi(\xi^\prime) $ and $\chi^2=\delta/2 - \sqrt{ \delta^2/4 -2}$.
Coefficient $C$ is  determined from the matching $W(x)$  with the 
asymptotic of one-dimensional interface solution.
Eq.(\ref{contour})  is similar to that of Ref. \cite{gold}. 
 The first term in the r.h.s. of
(\ref{contour}) at $\sigma>0$ describes the destabilizing effect of negative surface tension,
 and the Biot-Savart type integral accounts for non-local interactions among 
parts of the interface. 
According to Ref.\cite{gold}, combination of these effects gives rise to labyrinthine 
patterns. Since $\chi$ here is complex, function $K_0$ oscillates,
and interfaces form bound states at the distance $l\sim O(\chi^{-1}$). 
The characteristic wavelength of labyrintine patterns is  $l$ and is different 
from the most unstable wavelength of the original interface instability. This difference is evident 
from Fig. \ref{interf_fig} which shows snapshots of the  interface dynamics. 


In the region of stability ($\beta>-1$)  for the original model
Eq. (\ref{eq1}) the interface is stationary due to symmetry.
However, if the plate oscillates with two frequencies, $f$ and
$f/2$, the symmetry between two states connected by the interface, is
broken, and interface moves. The velocity of interface motion depends
on the relative phase of the sub-harmonic forcing with
respect to the forcing at $f $. This effect can be described by the
additional term  $q e^{i\Phi}$ in Eq. (\ref{eq1}), 
where $q$ characterizes the amplitude of the sub-harmonic pumping, and
$\Phi$ determines its relative phase.
For small $q$, we look for moving interface solution in the form
$\psi= \psi_0(x-vt) + q \psi_1(x-vt) + ... $. 
and $v=O(q)$.
Solvability condition yields the
following expression for the interface velocity
\begin{equation}
v=-q\frac{\cos\Phi \int{A^+dx} + \sin\Phi \int{B^+dx}}
{\int{(A^+\partial_xA_0+B^+\partial_xB_0)dx}}
\label{speed}
\end{equation}
The explicit answer is possible to obtain  for $b=0$  when $A^+=\partial_x A_0$, 
and $\Phi=0,\pi$, which yields 
the interface velocity  $v=\mp \frac{3}{2}q A_0^{-2}=\mp
 \frac{3}{2}q (s-1)^{-1}$. 
In general,  $A,B$, $A^+,B^+$, and hence $v$, can be 
found numerically.  The 
interface velocity as function of $q, \Phi$ is shown in Fig. \ref{interf_speed}.

We have shown that large acceleration  patterns  are captured by the generic 
parametric Ginzburg-Landau equation (\ref{eq1}).
The structure of the phase diagram Fig.\ref{phase} is qualitatively similar to that of 
experiments Refs.\cite{swin1,blair,umban} for high frequencies of vibration.
Increasing vibration amplitude leads to transition from a trivial state to stripes, hexagons, 
decorated interfaces, and finally, to stable interfaces. In experiments, at yet higher $\Gamma$, 
quarter-harmonic patterns appear, however these patterns are not described by our model.
Transition from unstable to stable interfaces also occurs with decreasing $\omega$ (increasing 
vibration frequency $f$), in agreement with Ref.\cite{blair,umban}. In our model, the interface instability
leads to labyrinthine patterns, however it seems that in experiments this instability sometimes saturates 
to provide stationary ``decorations''. Presumably, this saturation is caused 
by the weak interaction with the density field $\rho$ neglected here.
For additional sub-harmonic driving the model 
predicts steady moving interfaces with  the direction of  motion
controlled by the phase of the sub-harmonic component.
Experimental study of the controlled interface motion will be reported 
elsewhere \cite{exp}. 

We  acknowledge useful discussions with H. Swinney, 
R. Behringer, H. Jaeger and R. Goldstein. 
This work was supported by the US DOE  under grants
DE-FG03-95ER14516,  DE-FG03-96ER14592,  
 W-31-109-ENG-38,   and by NSF, STCS 
\#DMR91-20000. 
\leftline{\psfig{figure=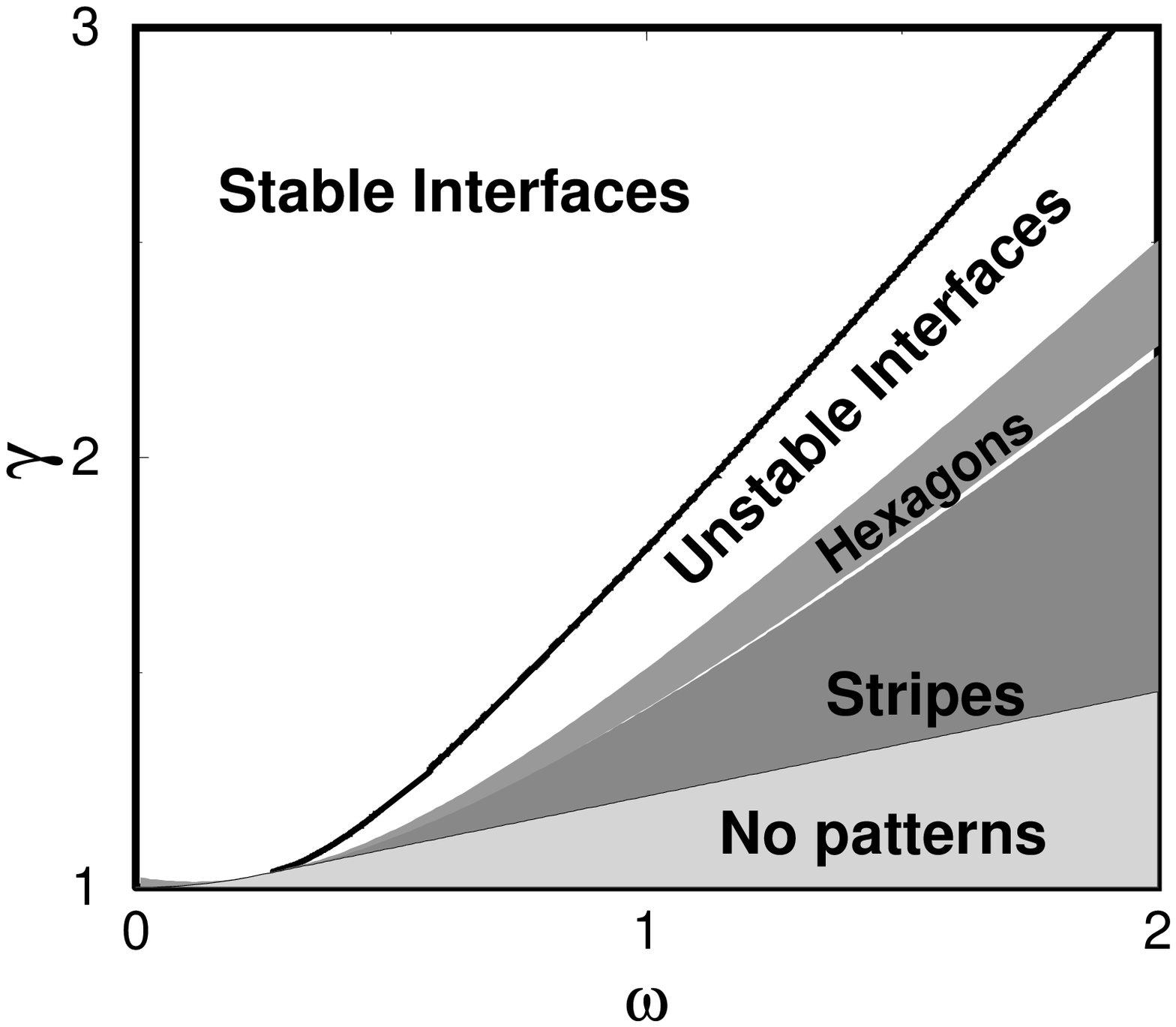,height=2.25in}}
\begin{figure}
\caption{Phase diagram for Eq. (\protect\ref{eq1}) at  $b=4$.}
\label{phase}
\end{figure}

\leftline{\psfig{figure=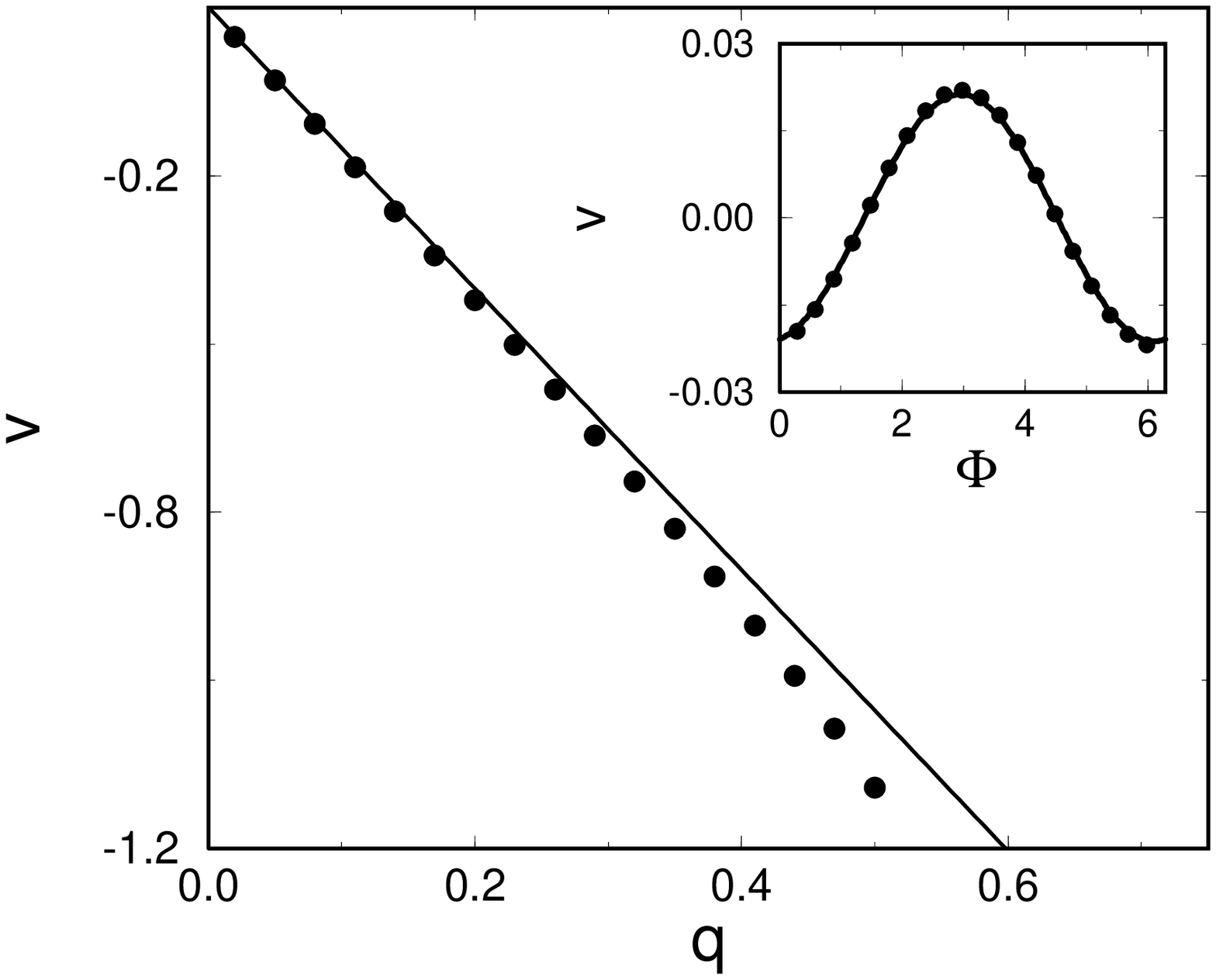,height=2.25in}}
\begin{figure}
\caption{Interface velocity $v$ for $\omega=1, b=4, \gamma=2.5$ 
vs $q$ at $\Phi=0$.
Inset: $v$ vs  $\Phi$ at  $q=0.01$. 
($\bullet$) - numerical results, (---) - analytical expression
(\protect \ref{speed}).  
}
\label{interf_speed}
\end{figure}
\end{document}